\documentclass{aastex}
\usepackage{emulateapj5}
\usepackage{apjfonts}

\newcommand{\rl}{R_{\rm L}}
\newcommand{\Mdot}{\dot{M}}
\newcommand{\Lsun}{L_{\odot}}
\newcommand{\Msun}{M_{\odot}}
\newcommand{\pz}{P_{\rm ZAMS}}
\newcommand{\poz}{P_{0}/\pz}
\newcommand{\be}{\begin{equation}}
\newcommand{\ee}{\end{equation}}

\begin{document}

\shorttitle{Case A Binaries}

\title{A Complete Survey of Case A Binary Evolution with Comparison to
Observed Algol-type Systems}

\author{
	C.A. Nelson\altaffilmark{1,2} \&
	P.P. Eggleton\altaffilmark{1,3}
        }

\altaffiltext{1}{Lawrence Livermore National Laboratory, Livermore, CA 94550\\
    Email: {\tt cailin@llnl.gov, ppe@igpp.ucllnl.org}}

\altaffiltext{2}{Department of Physics, University of California, Berkeley,
        CA 94720}

\altaffiltext{3}{On leave from the Institute of Astronomy, Madingley Rd,
         Cambridge CB3 0HA, UK}

\begin{abstract} 
We undertake a comparison of observed Algol-type binaries with a library of
computed Case A binary evolution tracks.  The library consists of 5500 binary
tracks with various values of initial primary mass $M_{10}$, mass ratio
$q_{0}$, and period $P_{0}$,  designed to sample the phase-space of Case A
binaries in the range $-0.10 \le \log M_{10} \le 1.7$. Each binary is evolved
using a standard code with the assumption that both total mass and orbital
angular momentum are conserved. This code follows the evolution of both stars
until the point where contact or reverse mass transfer occurs.  The resulting
binary tracks show a rich variety of behavior which we sort into several
subclasses of Case A and Case B.  We present the results of this
classification, the final mass ratio and the fraction of time spent in Roche
Lobe overflow for each binary system.  The conservative assumption under which
we created this library is expected to hold for a broad range of binaries,
where both components have spectra in the range G$0$ to B$1$ and luminosity
class III -- V.  We gather a list of relatively well-determined observed hot
Algol-type binaries meeting this criterion, as well as a list of cooler
Algol-type binaries where we expect significant dynamo-driven mass loss and
angular momentum loss.  We fit each observed binary to our library of tracks
using a $\chi^2$-minimizing procedure.  We find that the hot Algols display
overall acceptable $\chi^2$, confirming the conservative assumption, while the
cool Algols show much less acceptable $\chi^2$ suggesting the need for more
free parameters, such as mass and angular momentum loss. 
\end{abstract}


\section{Introduction}
Many binary stars are observed to be undergoing Roche-lobe overflow
(RLOF), which is recognised as being a natural response to the fact that,
for a binary of given separation, there is a critical maximum radius, the
Roche-lobe radius, that a star cannot exceed without losing mass to its 
companion. There are many sub-types of stars undergoing RLOF, but we 
concentrate here on those which, like the prototype Algol, consist of
(i) a lobe-filling, mass-losing star that is substantially above the
main sequence, and~(ii) a component which underfills its Roche lobe,
and is usually nearer to, though still larger than, the main sequence. 
We concentrate on those (Case A) with short initial periods,
the lower and upper period depending on the primary mass.

It is not difficult to evolve theoretically pairs of stars with a given
initial primary mass $M_{10}$, initial mass ratio $q_0$ and initial orbital
period $P_0$, and follow them into, and beyond, the stage of RLOF.  However,
such evolution is certainly affected by assumptions regarding both mass loss
and angular momentum loss from the system as a whole. As a zero-order model it
is commonly supposed that both total mass and orbital angular momentum are conserved, and we
have computed conservative evolution for a large number of binary initial
parameters: $37\times10\times 15$ models with various $M_{10}$, $q_0$ and
$P_0$. Most of the periods considered are appropriate to Case~A, but some
correspond to Case B.

There is plenty of evidence, both direct and indirect, that mass loss and/or
angular momentum loss takes place in at least some systems. If mass escapes
from the system as stellar wind, then it will also carry angular momentum
away. Mass loss is observed fairly directly both in cool stars, where it
appears to be driven by dynamo activity in their convective envelopes, and in
hot stars, where radiation pressure in spectral lines may be the main driving
force. Mass loss is also clearly evident in many stars of supergiant
luminosity, across the whole range of spectral types; but we do not consider
supergiants here. However there is a broad range of spectra, from about G0 to
perhaps B1 and luminosity class III -- V, where there is rather little
evidence of significant mass loss, and where the conservative assumption may
therefore be reasonable. We test this by comparing a selection of observed
`hot Algols' (having both spectra in this range) with theoretical conservative
models, using a $\chi^2$ test. We find a reasonable agreement, especially if
we exclude one system which is near the extreme of this temperature range.
Comparing the same conservative models against some observed `cool Algols' we
find, as we expect, that the agreement is much poorer.

We have used a massively-parallel array, the Compaq Teracluster 2000 at
LLNL, to evolve our data cube of models. This data cube covers the following 
ranges of initial primary mass $M_{10}$ (in solar units), initial mass 
ratio, defined by  
\be
q_0\equiv \frac{M_{10}}{M_{20}} >  1\ ,
\label{qdef}
\ee
and initial period $P_0$:
\begin{mathletters}
\begin{eqnarray}
\log M_{10}=-0.10, -0.05, \dots, 1.7\ , \\
\log q_0=0.05, 0.10,\dots,0.5\ , \\
\log(P_0/P_{ZAMS})=0.05,0.1,\dots, 0.75\ .
\end{eqnarray}
\label{grid}
\end{mathletters}
Here $P_{ZAMS}$, a function of $M_{10}$, is the period at which the initially 
more massive component would just fill its Roche lobe on the zero-age 
main sequence. We used the approximation

\be
P_{ZAMS} \approx \frac {0.19 M_{10}+0.47 M_{10}^{2.33}}{ 1+1.18 M_{10}^2}\ .
\label{pzams}
\ee

These initial periods cover Case A and a small part of Case B. 
We constructed such a `data cube' with each of six metallicities ($Z=0.03, 
0.02, 0.01, 0.004, 0.001, .0003$), and also, for $Z=0.02$ only, with three 
different assumptions about mass loss/angular momentum loss (in addition 
to the conservative assumption). We present here only the conservative, 
$Z=0.02$ data cube. 

In \S 2, we discuss the numerical modelling and the physical assumptions that
go into our data cube, and in \S 3 we discuss the results. We attempt to
classify the results into a small number of sub-categories of Case A (and some
analogues in Case B), depending for instance on whether the two components
come into contact rapidly, slowly, or not at all after the start of RLOF, and
(in the last case) on whether or not primary reaches a supernova before the
secondary swells up enough to reach reverse RLOF.  In \S 4 we discuss our
attempts to fit several observed semidetached systems (Algols) with the
theoretical models. We give our conclusions in \S 5.

We emphasise here that even if a particular Algol {\it can} be reasonably
fitted by a conservative model, this does not prove that the evolution was
conservative. Some models of non-conservation might lead to the same current
parameters, starting from different initial conditions.  Even if we had a mass
loss/angular momentum loss model with no free parameters in it, we might still
have ambiguity, partly because there are only six independent observational
parameters (current $P,~M_1,~M_2,~R_2,~T_1,~T_2$) to be fitted by four
theoretical parameters (age, and $P_0,~M_{10},~q_0$), and partly
because our data cube is still quite coarse even with 5550 models in it.

We also emphasise that throughout this paper we use suffixes 1 and 2
consistently to refer to the components with the greater and smaller
\textit{initial} mass respectively. This may seem unfortunate since observers
normally call the currently hotter (and normally more massive) component the
`primary', at least in Algol systems. This component is the descendant of the
originally less massive star. We do not think it would be helpful to
interchange the suffices at the points in evolution where the ordering of the
temperatures changes. However to avoid the most obvious possibility of
confusion we do not use the terms `primary' and `secondary': instead we refer
to the components as $*1$ (pronounced 'star one') and $*2$, and keep these
designations throughout their entire evolution. The mass ratio $q$ as defined
in Equation~\ref{qdef} always starts off with $q_0 > 1$. After some RLOF it is
commonly $< 1$.

\section{The Theoretical Data Cube}

 We used the stellar evolution code most recently described by Pols
et al. (1995), based on the code of \citet{egg71,egg72,egg73}. This code
is fully implicit in the composition equations as well as in the structure
and the mesh-spacing equations. The
implicit adaptive mesh is particularly useful for mass-transfer
situations. In fact, it means that in a first approximation we do not have 
to do anything to the code to account for mass transfer, except replace 
a boundary condition $M(t)={\rm constant}$ by a condition which gives the 
mass-loss rate $\Mdot$ as a function of stellar radius $R$ and Roche-lobe 
radius $\rl$ \citep{tou88}.

We will not repeat here a description of the physical input (Pols
et al. 1995). We have however included a simplistic model of convective
overshooting \citep{sch97,pol97}, based on a comparison of theoretical and 
observed {\it non}-interacting binaries. Other assumptions in the code are 
standard, and include the following: (a) the convective mixing of 
composition is treated as a diffusion equation, with diffusion coefficient 
a function of $\nabla_r-\nabla_a$ \citep{egg72}, and (b) because the mesh 
is fully adaptive, i.e. non-Lagrangian, an upstream advection term is needed 
in all time derivatives \citep{egg71}. The former ensures that any convection
zones satisfy the K. Schwarzschild convection criterion 
($\nabla_{\rm r}\approx\nabla_{\rm a}$) and simultaneously that any semiconvection zones that may 
arise satisfy the M. Schwarzschild condition \citep{sch58}
and are dealt with automatically, without extra code; the latter ensures that 
any evolutionary stage involving thin burning shells is computed very efficiently.

Regarding situations specific to binaries, we make the following 
assumptions:
\renewcommand{\theenumi}{\roman{enumi}}
\renewcommand{\labelenumi}{(\theenumi)}
\begin{enumerate}

\item The star is still treated as spherically symmetric, the radial
coordinate $r$ being the volume-radius of an equipotential surface. 
The gravity at effective radius $r$ is reduced by a factor dependent on angular 
velocity:

\begin{equation}
g = \frac{Gm}{r^2} \left(1-\frac{2\Omega^2r^3}{3Gm}\right)\ ,
\label{gdef}
\end{equation}

where $m$ is the mass within an equipotential of volume-radius $r$ and $\Omega$ 
is the angular velocity of the star, assumed corotating with the binary.
\def\rl{R_{\rm L}}
\item Mass transfer from a star that overfills its Roche lobe is treated as 
spherically symmetric, and governed by the boundary condition
\begin{equation}
\begin{array}{cc}
\Mdot & =    -C \{ \log(R/\rl)^{3}\},\ \ R>\rl \\
 & =   0,\ \  R <\rl
\end{array}
\label{mdot}
\end{equation}
\def\th{\thinspace}
where $C=500\th\Msun/$yr. Thus a transfer rate of $5\times 10^{-7}\Msun/$yr
corresponds to an overfill of 0.1\%.
We only do this for $*1$. There inevitably comes a point in evolution
when $*2$ fills its own Roche lobe, but this is usually either
(a) while $*1$ already fills its own lobe, so that the binary comes into
contact -- for the present, we stop evolution at this point; or
(b) when $*1$ has evolved to a late and relatively compact state of low
mass, and $*2$ has grown to a very large radius. In the latter case the mass 
ratio is very small ($q \lesssim 0.2$). Therefore the mass transfer can be 
expected to be rapid, and unstable on a short (hydrodynamical) timescale. We
expect common-envelope evolution beyond this point \citep{pac76},
and so we stop evolution at this point, also.

\item It is assumed that the matter which leaves $*1$ is accreted in a
spherically symmetric manner at the surface of $*2$, with entropy and
temperature equal to the surface values of $*2$. Thus no model is 
incorporated for the temperature/entropy budget of the material during transfer.
This may seem potentially serious, but when most of the mass is transferred
on a nuclear timescale it should not be important. 

\item The composition of accreting material on $*2$ is assumed to be 
the same as that of material already just below the surface of $*2$, rather than 
(as it should be) of the material leaving $*1$. This is just done for
convenience, and is only significantly in error at a fairly late stage
in mass transfer. The observed Algols that we make comparison with are
probably not at such late stages.

\item On a somewhat technical level, the implementation of
Equations~\ref{gdef} and ~\ref{mdot} numerically within the framework of a
fully implicit and adaptive code means that it is desirable, though possibly
not essential, to introduce an extra equation into the usual set of difference
equations -- for the structure, composition and mesh-distribution variables --
that are solved for by Newton-Raphson iteration. This is because Equation (4),
while depending primarily on the local variables $r$ and $m(r)$, also depends on the
{\it surface} mass $M(t)$, through $\Omega(t)$. $\Omega$ depends not only on
the orbital angular momentum (which, being assumed constant in a conservative
model, is no problem) but also on the the masses of the two components via
Newtonian gravitation. Because of Equation (5), $M(t)$ is not known {\it a
priori}, but only after the iteration is finished. We found it convenient to
add $M^{\prime}(t,r)$ as a new but somewhat artificial variable satisfying the
trivial equation
\be
\frac{\partial M^{\prime}}{\partial r} =0~,
\ee
with the equally trivial boundary condition that at the surface
\be
M^{\prime}(t,r) = M(t)\ .
\ee

Although this modification is barely necessary for the conservative
models, it is rather more important for non-conservative models, where
Equation~\ref{mdot}  may have an extra term, attributable to stellar wind, and where
the angular momentum is no longer constant.
\end{enumerate}

With the above assumptions and modifications the code works reasonably
satisfactorily in an automatic way. We set up a grid of starting models with
$M_{10}$, $\log q_0$ and $P_0$ given by Equation~\ref{grid}.  For most masses in the
range $1.5\Msun \leq M\leq 16\Msun$ we found that we obtained Case A evolution
with $1 < \poz \lesssim 4$, and Case B for $\gtrsim 4$, but the critical value
for Case B decreased rapidly below $\sim 1.5\Msun$, and increased slowly above
$\sim 16\Msun$.

Given $M_{10}$, $q_0$ and $P_0$, we started by evolving $*1$ until one 
or other of the following conditions occurred:
\renewcommand{\theenumi}{\alph{enumi}}
\renewcommand{\labelenumi}{(\theenumi)}
\begin{enumerate}
\item 2000 timesteps were taken
\item carbon-burning luminosity exceeded $1\Lsun$, indicating that a
supernova explosion was imminent
\item the age exceeded $20$ Gyr
\item the code failed to converge, or
\item the stellar radius exceeded the Roche-lobe radius by more than $10\%$.
\end{enumerate}

 For $*1$, (e) indicated that hydrodynamically unstable RLOF was taking place, usually 
due to a large initial mass ratio ($q_0 \gtrsim 3$), or to a deep convective
envelope on the loser.

We then ran $*2$, giving it a rate of mass gain that was the negative
of the stored mass loss rate of $*1$. This run was also terminated at the first
point when one of the first four conditions above occurred, but it could also
terminate itself if

\renewcommand{\labelenumi}{(f)}
\begin{enumerate}
\item the age of $*2$ went beyond the age at which $*1$ terminated, or
\end{enumerate}
\renewcommand{\labelenumi}{(g)}
\begin{enumerate}
\item the radius of $*2$ reached its Roche-lobe radius.
The latter normally meant either that the system had 
evolved into contact, $*2$ filling its lobe while $*1$ still was, or else
that it had evolved into a reverse RLOF situation, with $q < 0.2$,
that would presumably lead to mass transfer on a hydrodynamical timescale,
probably implying common-envelope evolution. In either case, the implicit
assumption that $*1$'s evolution is independent of whatever happens to $*2$
breaks down, and so we consider here only the evolution that takes place
prior to the point where $*2$ filled its Roche lobe.
\end{enumerate}

Convergence failure -- (d) above --  was not very common, though more common
than we would have wished. For $*1$ it was usually because either (i)
Equation~\ref{pzams} apparently gives slightly too small a value, for some
ranges of $M_{10}$, so that at the lowest value of $P_0$ for those masses $*1$
already filled its Roche lobe while still making a rapid adjustment from the
approximate ZAMS from which it started; or (ii) for the most massive stars,
$\gtrsim 25 \Msun$, a breakdown often occurred when $*1$ approached a sloping
line across the HRD, starting just before the terminal main sequence at $\sim
50\Msun$ (our highest initial value) and reaching to the red supergiant region
at $\sim 25\Msun$. It may not be just coincidence that this is also
approximately the observational `Humphreys-Davidson Limit', which appears to
be an upper limit for stars in the HRD. Stars close to this limit are
typically P Cyg stars, Hubble-Sandage variables, or Luminous Blue Variables
(LBVs). Such stars have internal luminosities that are close to or even above
the Eddington limit in zones where the opacity has a local maximum. Thus it
may be that the numerical convergence difficulties have their origin in the
physical difficulty of maintaining hydrostatic equilibrium in such stars.

For $*2$ convergence failure occurs because:
\renewcommand{\theenumi}{\roman{enumi}}
\renewcommand{\labelenumi}{(\theenumi)}
\begin{enumerate}
\item  In binaries with extreme initial mass
ratios $q_{0} \gtrsim 4$, $*2$ often fails to converge while gaining mass at
the thermal timescale of $*1$. This may occur because the large mass ratio
means that the thermal timescale of $*1$ is closer to the dynamic
timescale of $*2$ 
\item In our lowest mass binaries, $M_{20} \lesssim 0.80
\Msun$, the mass gaining star ($*2$) `ignores' the fact that $*1$ is losing
mass for a handful of timesteps and maintains a constant mass.  Then $*2$ 
attempts to gain all the mass that $*1$ has lost in of order ten timesteps in
a single timestep and fails to converge
\item  When the mass gaining star
has a mass $\sim 1.5 \Msun$.  These stars are in the transition region between
lower MS stars with convective envelopes and upper MS stars with convective
cores, and they possess very shallow surface convection zones which may only
be a few mesh points wide.  We suspect that this barely resolved surface
convection zone contributes to their numerical instability; or, finally, 
\item we also see a theoretical `Humphreys-Davidson Limit' in $*2$ at high mass.
\end{enumerate}

\section{Classification of Types of Evolution}
\begin{figure}
\epsscale{0.5}
\plotone{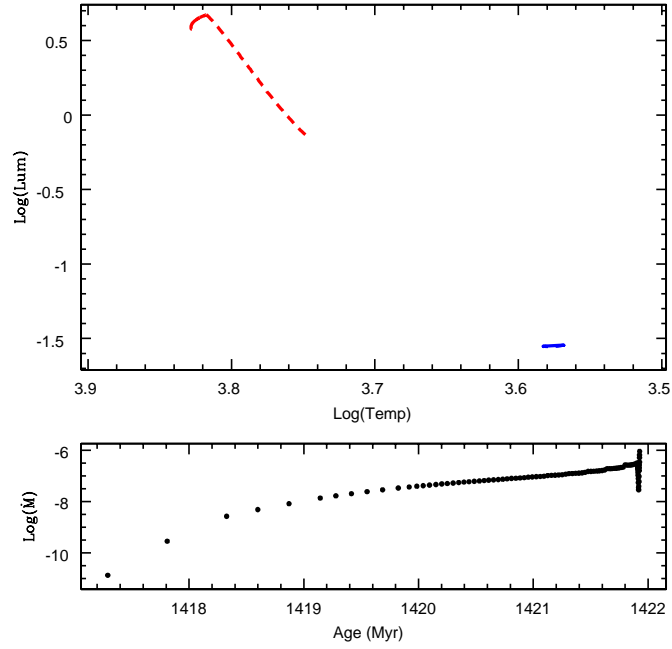}
\caption{Case AD - Dynamic Contact. In the top panel of this figure, and in
Figures ~\ref{ar} - \ref{an}, we show the evolution of both stars in an HR
diagram.  The track for $*1$ is always the initially more luminous, $*2$ the
initially less luminous.  Solid lines indicate periods where the binary is
separated, dashed lines indicate periods where $*1$ is transferring mass to
$*2$.  We mark any transitions to the Hertzsprung Gap as H, and transitions to
the Giant Branch as G. The bottom panel shows the mass transfer rate in
logarithmic units of $\Mdot/ \rm $yr for the period in time during which mass
transfer occurs.  The initial parameters of this binary are $\log{M_{10}} =
0.15$, $q_{0} = 0.50$ and $\log\poz =0.15 $.  The mass transfer rises rapidly
to the dynamic timescale and the two stars come into contact. }
\label{ad}
\end{figure}

\begin{figure}
\epsscale{0.5}
\plotone{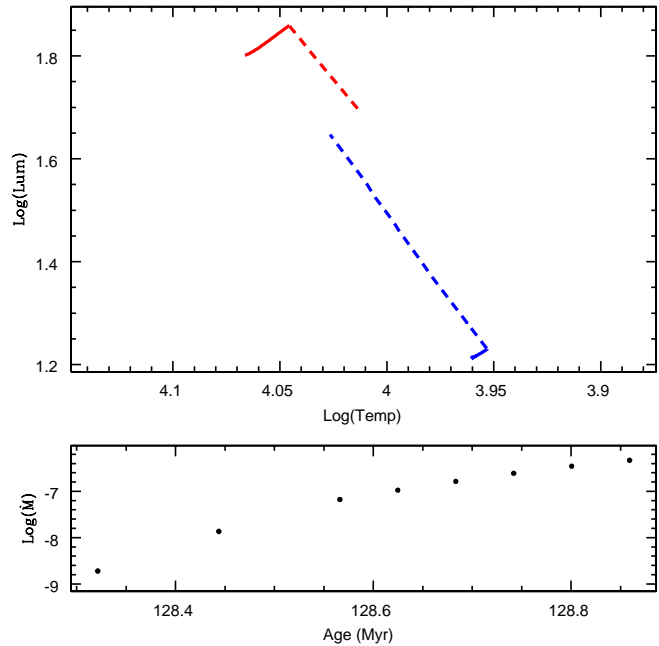}
\caption{Case AR - Rapid Contact.  The initial parameters of this binary are
$\log{M_{10}} = 0.45 $, $q_{0} = 0.15$ and $\log\poz = 0.10$. The mass
transfer rate rises rapidly to the thermal timescale and the two stars come
into contact with a final mass ratio $\log q = 0.11$, still well above unity. }
\label{ar}
\end{figure}

We define here the six major subtypes of Case A evolution identified by
\citet{egg00}, cases AD, AR, AS, AE, AL, AN.  In addition, we define two
rather more rare cases, AG and AB.  Three of these subtypes (AD, AR, AS) lead
to contact while both components are on the main sequence (MS).  Two cases
(AE, AG) reach contact with one or both components evolved past the terminal
MS.  After the initial episode of mass transfer from $*1$ to $*2$, the
remaining three cases experience a period of separation followed either by
reverse mass transfer at very small $q$ (AB, AL) or the supernova of $*1$
(AN). Specifically, the six cases, are:

\begin{itemize}

\item AD -- dynamic RLOF: this occurs in binaries with large $q_0$ and in
binaries where the mass losing star ($*1$) has a deep convective envelope.
Once RLOF begins, mass transfer quickly accelerates to the dynamic timescale
of $*1$, $t_{\rm dyn}$,  which we assume to be less than a tenth of the thermal timescale,
$t_{KH}$.  The thermal (or Kelvin-Helmholtz) timescale is determined in the 
code as the integrated total energy, thermal plus gravitational, divided by 
the total luminosity at the surface. Thus the mass transfer is determined to 
be dynamic when

\be
\Mdot > 10.0 \times \frac{M}{t_{\rm KH}}\ .
\ee

The calculation is terminated by (e) above, but seems likely to lead either
to contact or to a common-envelope situation, and probably then to a complete
merger of the two components.  We illustrate the behaviour in the HRD and
the mass transfer rate of case AD in Figure~\ref{ad}.

\item AR -- rapid evolution to contact: this occurs in binaries with moderate
to large $q_{0}$. In these cases $*2$ expands so rapidly in response to the
onset of $*1$'s thermal-timescale RLOF that it fills its own Roche lobe before
much mass is transferred. We define the mass transfer rate to be thermal when
the magnitude of the thermal luminosity, $\mid L_{\rm therm} \mid$, reaches
2\% of the nuclear burning luminosity, $L_{\rm nuc}$. This probably leads to a
contact binary of the W UMa type, although it can happen as easily for massive
stars (provided $q_0$ is suitably large) as for the lower masses of typical W
UMa systems.  Case AR behavior is illustrated in Figure~\ref{ar}.

\end{itemize}

In some binary runs, these two cases are difficult to distinguish.  While
evolution of $*1$ will proceed through several timesteps of dynamic timescale
mass transfer before being terminated by (e), the calculation of $*2$ is often
unable to converge while gaining mass at this rate.  The calculations of case
AR and AD binaries at very large $q_{0}$, therefore, often terminate before
contact is reached and we must guess the rate mass transfer achieves
before contact occurs.

To do so we extrapolate the function $\log { \frac{R_{2}}{RL_{2}}(t) }$ to the
time $t_{\rm contact}$ at which the radius of $*2$ has expanded to fill its RL
and $\log { \frac{R_{2}}{RL_{2}}(t) } = 0$.  We then examine the mass loss
history of $*1$ (whose calculation has proceeded further in time than that of
$*2$) and determine whether the mass transfer rate reaches the thermal or
dynamic timescale at $t \leq t_{\rm contact}$.  Unfortunately, the function
$\log { \frac{R_{2}}{RL_{2}}(t) }$ can be both non-linear and slightly noisy,
and so $t_{\rm contact}$, along with the maximum mass transfer rate, can
depend rather sensitively on the exact point in time at which $*2$ 
fails.

\begin{figure}
\epsscale{0.5}
\plotone{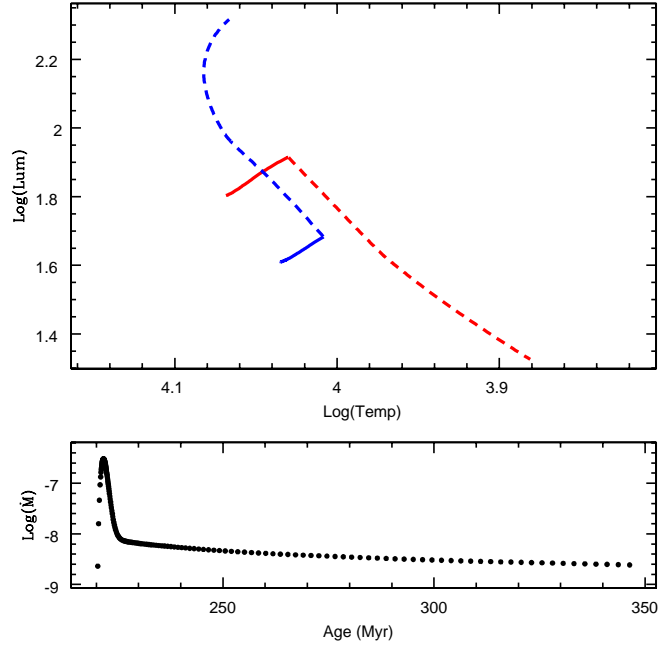}
\caption{Case AS - Slow Contact. The initial parameters of this binary are
$\log{M_{10}} = 0.45$, $\log{q_{0}} = 0.05 $ and $\log\poz = 0.20$. Contact
is avoided during the period of thermal scale mass transfer and a long
period of nuclear timescale mass transfer follows.  The run ends in contact
with a final mass ratio, $\log{q}= -0.27$.}
\label{as}
\end{figure}

\begin{figure}
\epsscale{0.5}
\plotone{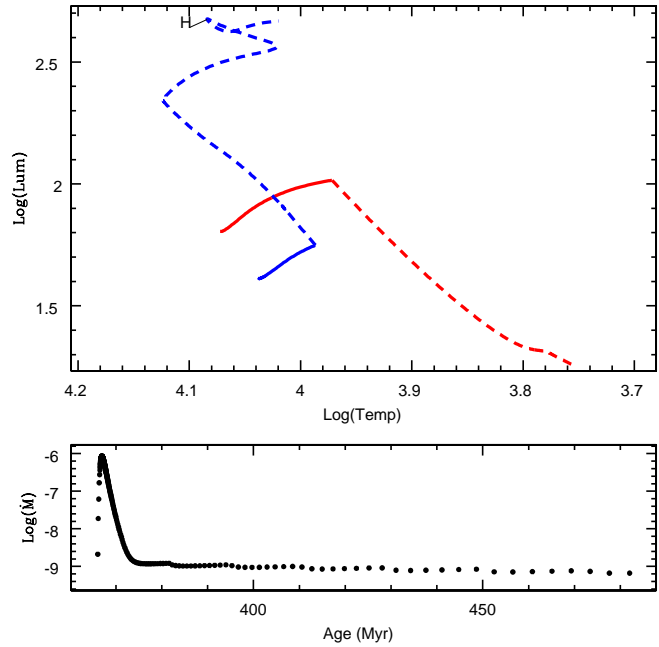}
\caption{Case AE - Early Overtaking. The initial parameters of this binary are
$\log{M_{10}} = 0.45$, $\log{q_{0}} = 0.05 $ and $\log\poz = 0.45$.  $*2$
gains so much mass that its evolution overtakes that of $*1$  and $*2$
reaches the HG first.  The run ends in contact with a final mass ratio of
$\log{q}= -0.23$.} \label{ae}
\end{figure}

\begin{itemize}
\item AS -- slow evolution to contact: this occurs in binaries with small $q_{0}$ and
small $P_{0}$.  These binaries experience a short burst of thermal timescale
mass transfer, followed by a long phase of nuclear timescale mass transfer,
during which much mass is exchanged.  The two stars come into contact slowly,
but reach contact before either star has left the MS.  The large amount
of mass transfer leads to a final mass ratio substantially below unity
(typically $q \sim 0.4 - 0.6$), and with both stars substantially larger than their
ZAMS radii.  Case AS behavior is illustrated in Figure~\ref{as}.  We note that
while $*2$ always remains near the main-sequence band, $*1$ evolves to
substantially cooler temperatures.  This is a common configuration in observed
Algol systems.

\item AE -- early overtaking: this occurs in binaries with small $q_{0}$ and
moderate $P_{0}$.  It occurs only in binaries with initial masses $2 \Msun
\lesssim M_{1} \lesssim 10 \Msun$.  The mass transfer in this case is very
similar to case AS.  In case AE, however, $*2$ gains so much mass that its
evolution is accelerated to the extent that $*2$ reaches the Hertzsprung Gap,
HG,  while $*1$ is still on the MS; the evolution of the initially less
massive star, $*2$, has {\it overtaken} that of $*1$.  We define the
overtaking as {\it early} because it occurs with $*1$ still on the MS. Most
case AE binaries reach contact shortly thereafter. However, in a few cases
$*1$ shrinks very slightly inside its RL at the end of the calculation and the
run ends with the RLOF of $*2$.  In these cases, $*1$ has very nearly
exhausted hydrogen and it is likely that it will soon swell once again to fill its RL and
contact will again occur.  Case AE behavior is illustrated in Figure~\ref{ae}.
\end{itemize}

In most cases where contact is avoided while $*1$ is on the MS, $*1$ loses so
much mass that it eventually shrinks inside its RL leaving only a compact
core.  A period of separation ensues which may then be followed by further
RLOF of $*1$ or $*2$.  These are the cases AL, AB and AN, described in more
detail below.  However, in our lower mass binaries ($M_{10} \leq 1.6 \Msun$)
we see a few cases where contact is avoided while $*1$ is on the MS, but
reached later on. 

\begin{figure}
\epsscale{0.5}
\plotone{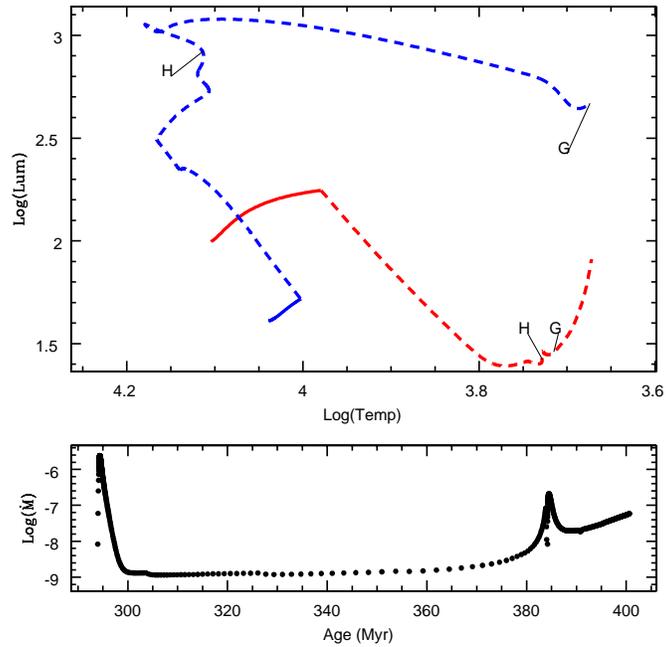}
\caption{Case AG - Contact on Giant Branch.  The initial parameters of this
binary are $\log{M_{10}} = 0.50$, $\log{q_{0}} = 0.10 $ and $\log\poz = 0.55$.
The stars come into contact at a mass ratio of $\log{q} = -0.83$.  There is a
brief period of separation at $t \sim 384$ Myr with $*1$ in the HG.}
\label{ag}
\end{figure}

\begin{itemize}
\item AG -- contact on giant branch: this occurs for $M_{10} \lesssim 1.6
\Msun$, and $P_{0}$ larger then those of AS/AE, but smaller then AL/AN.
Contact is avoided while $*1$ is on the MS, but occurs when $*1$
reaches the giant branch, GB.  At time of contact $*2$ is in the HG
or on the GB as well.  A typical example of case AG is shown in
Figure~\ref{ag}.
\end{itemize}

Cases AL, AN are distinguished by whether or not $*1$ supernovas before $*2$
reaches RLOF.  In practice, we assume a supernovae explosion to be iminent
when $*1$ begins burning carbon.

\begin{figure}
\epsscale{0.5}
\plotone{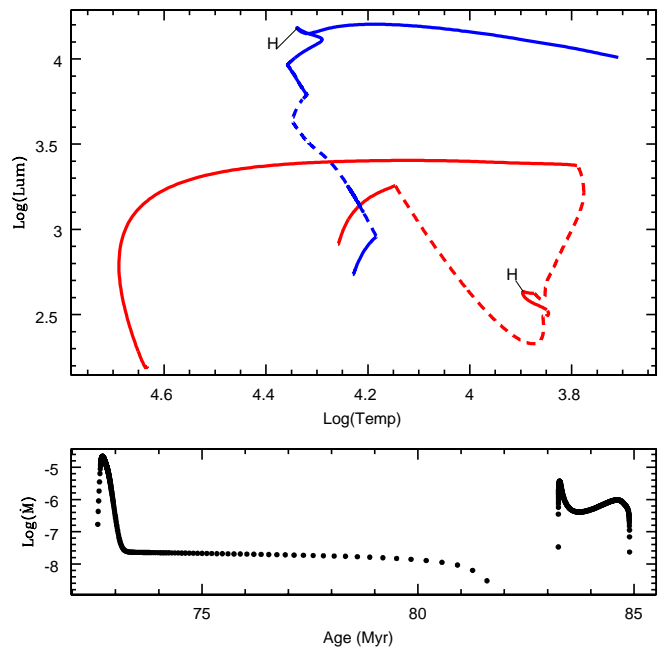}
\caption{Case AL - Late Overtaking. The initial parameters of this binary are
$\log{M_{10}} = 0.75$, $\log{q_{0}} = 0.05 $ and $\log\poz = 0.60$.  In this
run $*1$ loses so much mass in two periods of mass transfer that it eventually
shrinks inside its RL and becomes a low mass helium burning core, $\log{M_{1}} =
-0.07$.  The run ends as $*2$  crosses the HG and fills its RL at a very low
mass ratio, $\log q = -1.07$. The brief period of separation between the two
bursts of mass transfer is a feature common to all of our higher mass case
AL/AN binaries.  This feature occurs as $*1$ exhausts hydrogen in the core,
convection in the core shuts off and $*1$ shrinks slightly inside its RL. At
this point $*1$ behaves as a ``normal'' massive terminal MS star, executing
the classic hook at the end of the MS.  When hydrogen is completely exhausted
in the core and hydrogen shell burning begins $*1$ starts to cross the HG.
During this rapid phase of envelope expansion, $*1$ quickly fills its RL again
and mass transfer begins again.}
\label{al}
\end{figure}

\begin{figure}
\epsscale{0.5}
\plotone{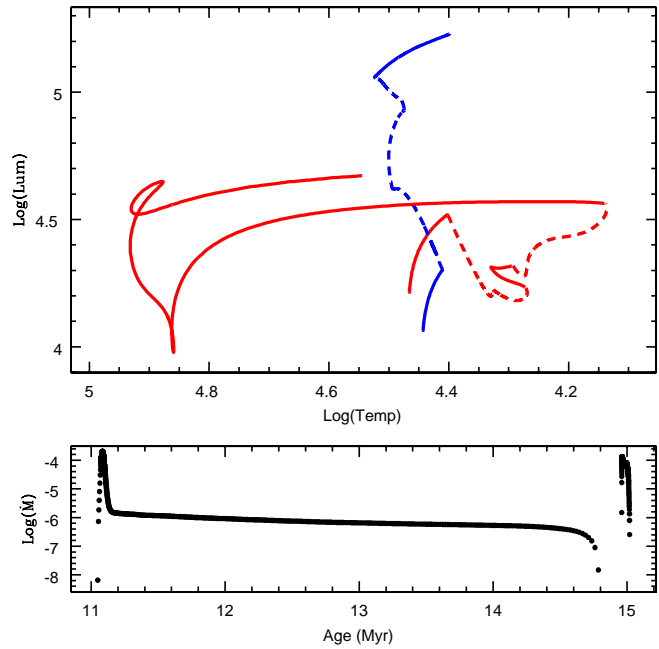}
\caption{Case AN - No Overtaking. The initial parameters of this binary are
$\log{M_{10}} = 1.15$, $\log{q_{0}} = 0.05 $ and $\log\poz = 0.45$.  After
two periods of mass transfer, $*1$ becomes a helium star of mass 
$\log {M_{1}} =  0.48$.  As $*2$ evolves towards the terminal MS,
$*1$ ignites carbon in the core. The ignition of carbon suggests an
imminent supernovae explosion, and we conclude that $*1$ will supernovae
before $*2$ fills its RL.}
\label{an}
\end{figure}

\begin{itemize}
\item AL -- late overtaking: this occurs in binaries with $M_{10} \lesssim 13
\Msun$ and moderate to large $P_{0}$.  In these binaries, $*2$ reaches RLOF
before $*1$ begins burning carbon. In many of the lower-mass AL cases, $*1$
has become a low mass remnant (WD or NS) which will never supernovae unless
the (uncomputed) reverse mass transfer results in significant mass gain for
$*1$. The evolution of $*2$ has {\it overtaken} the evolution of $*1$ in the
sense that the initially more massive star is now shrunk inside its RL while
the initially less massive star is undergoing RLOF.  The overtaking is {\it
late} because it occurs with $*1$ past the MS.  Case AL behavior is
illustrated in Figure~\ref{al}.

\item AN -- no overtaking: this occurs in higher mass binaries with moderate 
to large $P_{0}$.  In these binaries $*1$ reaches carbon burning,
indicating an imminent supernova, before $*2$ has reached RLOF.  Case AN
behavior is illustrated in Figure~\ref{an}.
\end{itemize}

\begin{figure}
\epsscale{1.0}
\plotone{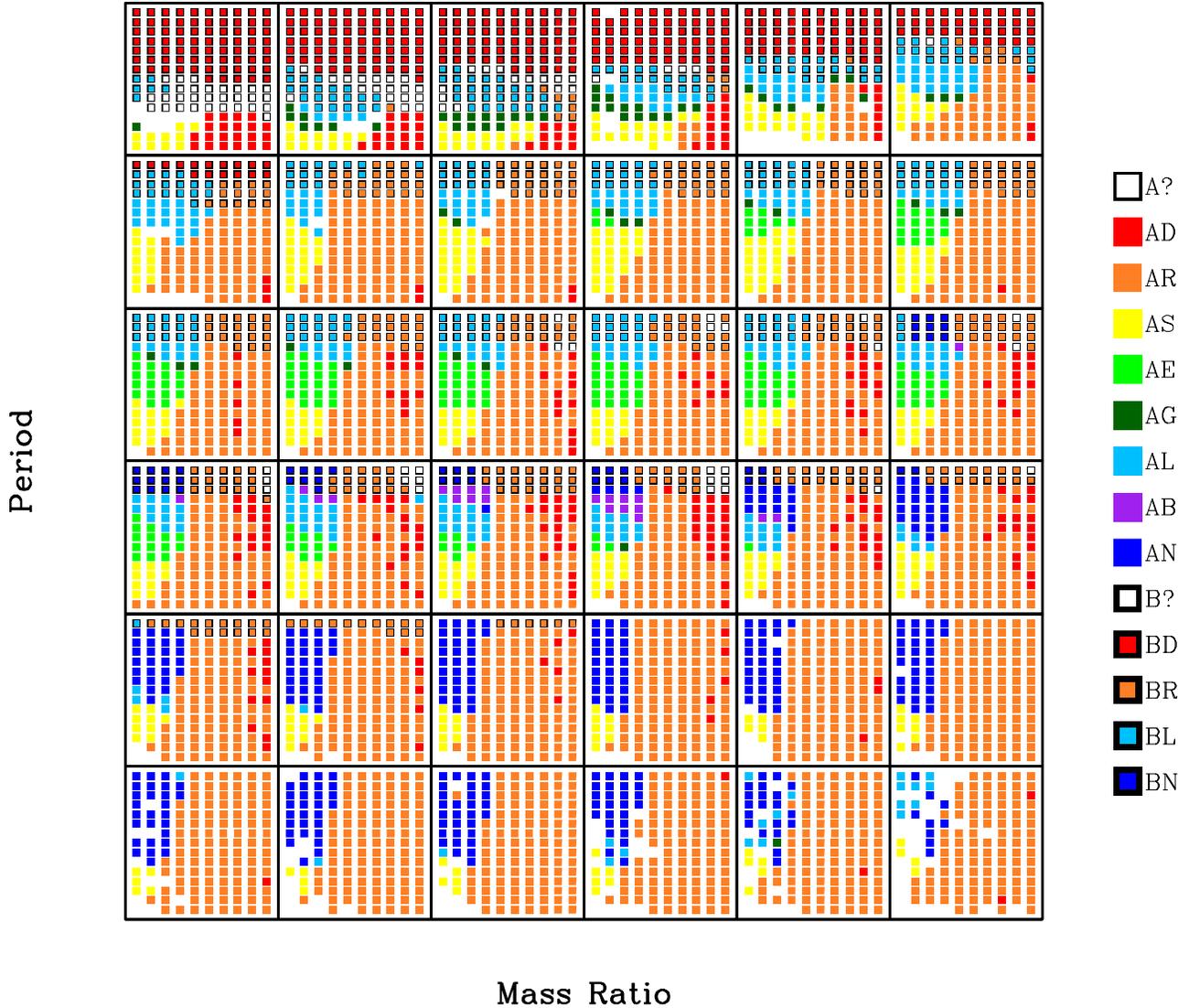}
\caption{The classification of each of our individual binary runs into cases
AD, AR, AS, AE, AG, AL, AB, AN, BD, BR, BL, and BN, as described in \S 3.
Each square represents a slice through the data cube at constant $M_{10}$.
The upper left hand block is a slice at $\log{M_{10}} = -0.05$.  The squares
increase in the order one reads the page of a book, increasing in units of
$0.05$ and ending in the bottom right hand corner with $\log{M_{10}}=1.70$.
We do not show our results for $\log{M_{10}} = -0.10$ as very few of these
binaries reached RLOF at an age younger then 20 Gyr.  Within each square the
x-axis represents increasing mass ratio in logarithmic units of $0.05$ from
$\log{q}=0.05$ to $0.50$.  The y-axis represents increasing period in units of
$0.05$ from $\log\poz = 0.05$ to $0.75$, where $\pz$ is the critical period at
which RLOF would occur on the ZAMS.  The color of each dot represents the
classification of the binary run according to the legend to the right of the
plot.  A white dot represents a Case A binary for which we could not determine
a sub-class, a white dot outlined in black indicates a Case B binary for which
we could not determine a sub-class.}
\label{state}
\end{figure}

As discussed in \S 2 the evolution of our most massive stars, $M_{10} \gtrsim 25
\Msun$ often breaks down.  This leads us again to the situation where we must
make a best guess as to what happens after the run stops.  To distinguish case AL
from AN we must determine whether $*2$ reaches RLOF before $*1$ ignites carbon.
This procedure is somewhat uncertain and leads to the great majority of our
unclassified runs at very high mass.  We also suspect that several of the
highest mass runs, $M_{10} \gtrsim 40 \Msun$, which were classified as AL are
uncertain, and may more probably be Case AN. (See Figure~\ref{state}.)

We note that the definitions of case AL and AN given here do not correspond
exactly to those of \citet{egg00}.  In this previous work, case AN also
included those binaries where $*1$ had become a WD or NS before $*2$ filled
its RL.  In this work, those binaries are included in case AL.

Pols (1994) noted that occasionally, in what we call case AL here, $*2$
could get to carbon ignition {\it before} $*1$, and thus be the first component
to explode as a supernova. \citet{pol94} modeled, in a simple way, the effect of the
presumed common-envelope phase in ejecting $*2$'s envelope, and then continued 
the evolution of the core. Although it would always have started He
burning later than $*1$, it might be sufficiently more massive to
overtake and ignite carbon first. However, in our work we did not attempt
to model the common-envelope phase at all, and so we cannot be definitive
about this possibility.

In addition, we include one more class: the classic Case AB. In our context
this is a subclass of case AL, where $*1$, after becoming a compact helium
core with a mass of $\sim 1 - 2\Msun$, expands again and experiences a further
period of RLOF.

\begin{figure}
\epsscale{0.5}
\plotone{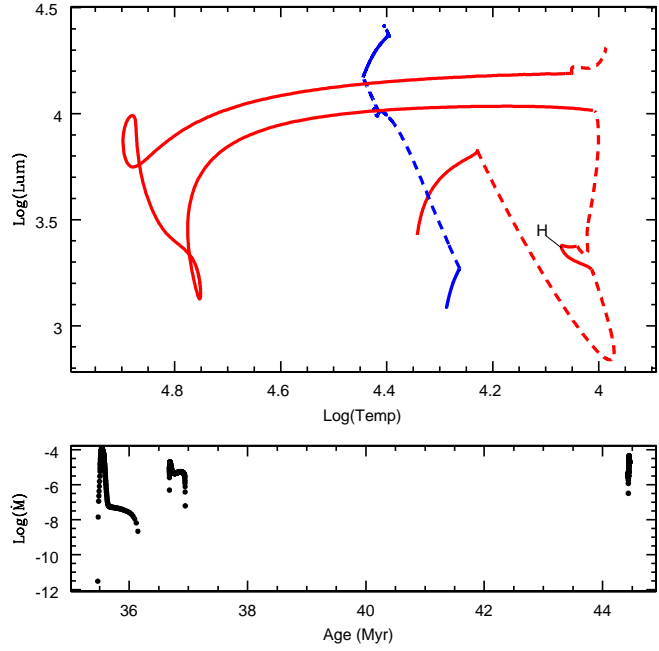}
\caption{Case AB - (Sub-type of AL).  The initial parameters of this binary are
$\log{M_{10}} = 0.90$, $\log{q_{0}} = 0.10 $ and $\log\poz = 0.65$.  
After igniting helium towards the end of the second burst of mass transfer,
$*1$ shrinks inside its RL for awhile, but eventually re-expands and undergoes
a third period of mass transfer.}
\label{ab}
\end{figure}

\begin{itemize}
\item AB - this occurs in binaries with $6 \Msun \lesssim M_{10} \lesssim 11
\Msun$, at small mass ratios and in a narrow range of periods between cases AL
and AN.  During the second burst of mass transfer, $*1$ ignites helium.  It
shrinks inside its RL for awhile, becoming a compact helium star.  It then
expands again and experiences a third period of mass transfer.  Although these 
binaries often
fail to converge at some point during this third period of mass transfer, we
suspect that it is followed by a period of separation and then reverse mass
transfer, making this a subclass of AL rather then AN. An example of case AB
evolution is shown in Figure~\ref{ab}.
\end{itemize}

In Table 1 we summarize the seven major subcases (excluding case AB),
providing the defining equations as well the evolutionary state and
geometrical configuration of the binary components at the end of the
calculation.  In this table we denote the main sequence as M, the Hertzsprung
Gap as H, the giant branch as G, and low and high mass remnants as R and C,
respectively.  In addition, we define the time of first RLOF, $t_{\rm RLOF}$,
the approximate MS lifetime of a single star, $t_{\rm MS}$, the time at which
the star enters the Hertzprung Gap, $t_{\rm H}$ and the time at which carbon
is ignited $t_{\rm Cburn}$.  We emphasize that we execute the classification
of each binary in our library automatically, and while the various clauses we
define work for the great majority of systems, we inevitably make a few
misclassifications.

As mentioned above, figures~\ref{ad} -~\ref{an}  illustrate the behaviour in
the HRD of the subtypes of Case A.  We also show the mass transfer rate for
times when $\Mdot > 0$.  Figure~\ref{state} shows which elements of our data
cube reached which outcome. Some of the systems of longer $P_0$ are Case B
rather than Case A.  These are usually analogous to either AD, AR, AL or AN.
Case BD is effectively the classical Late Case~B, where $*1$ reaches the giant
branch and acquires a deep convective envelope before RLOF begins; however it
can also be an extreme initial mass ratio rather than a convective envelope
which triggers dynamic mass transfer.  Case B systems, or at least those which
we have computed here, normally have fewer options than Case A because it is
difficult for $*2$ to catch up with $*1$ when $*1$ has already reached the
terminal main sequence before RLOF.  However, as emphasised by \citet{deg90},
it {\it is} possible for early Case B systems to show what we call here Case
BL for late overtaking, with $*2$ evolving to fill its own Roche lobe while
$*1$ has shrunk inside its own.  This kind of behaviour is particularly
prevalent in the mass range $M_{10} \lesssim 8\Msun$.

\begin{figure}
\epsscale{1.0}
\plotone{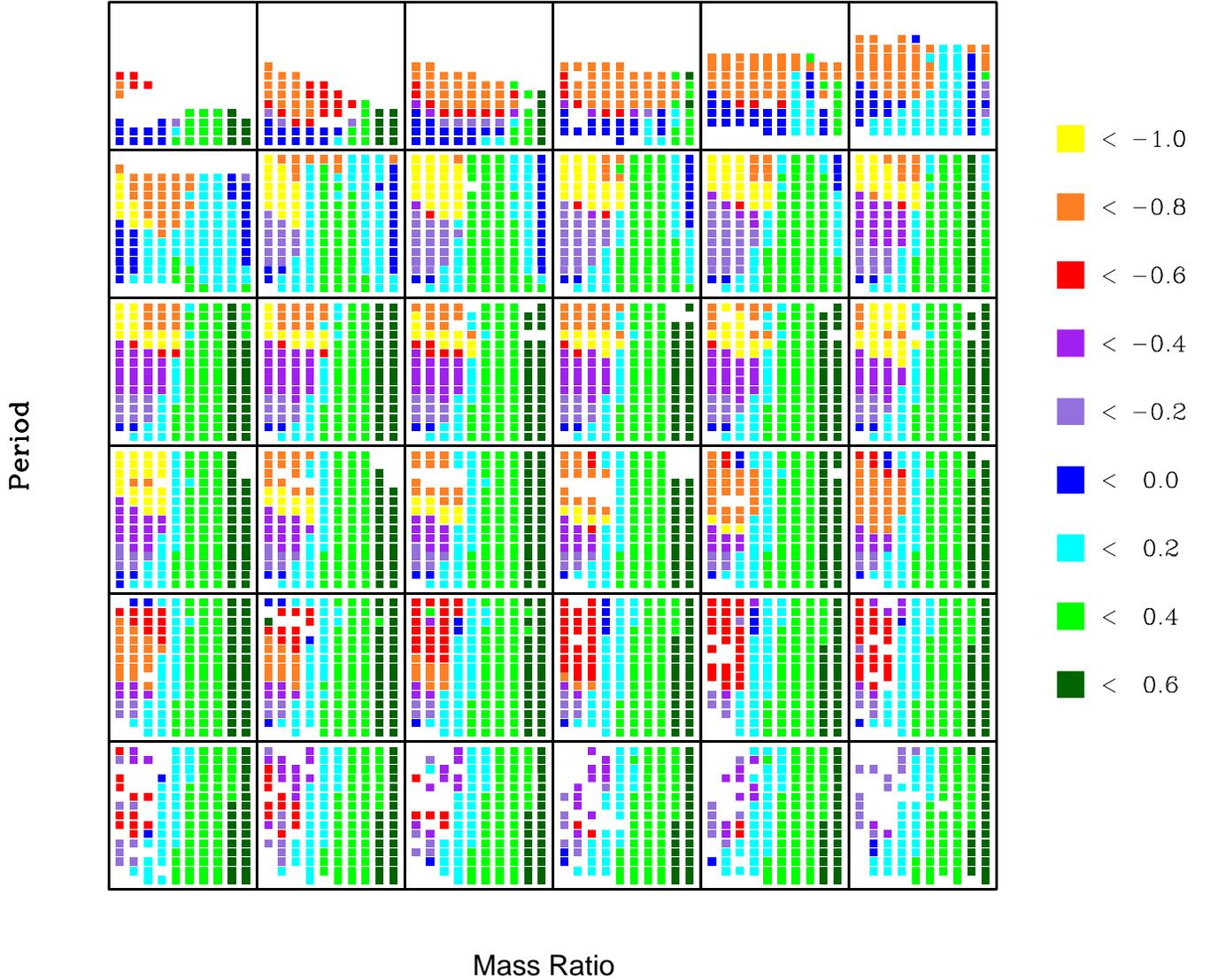}
\caption{The final mass ratio of each binary system. The
ordinates are defined as in Figure~\ref{state}. We define the final state to
be the point where either a) contact is reached or seems imminent, b) reverse
mass transfer begins, c) $*1$ has ignited carbon or d) the binary is in a
detached phase and reverse mass transfer is imininent. Binaries for which the
final state could not be determined appear as white squares. Higher mass white
squares are generally due to binaries which break down during thermal
timescale mass transfer.  The color-coded final mass ratios are in
logarithmic units.  The legend is to be interpreted as follows: 
binaries with $\log q < -1.0$ are shown as yellow
squares, those with $-1.0 \leq \log q < -0.8$ as orange squares,
and so forth.} 
\label{mratio}
\end{figure}

\begin{figure}
\epsscale{1.0}
\plotone{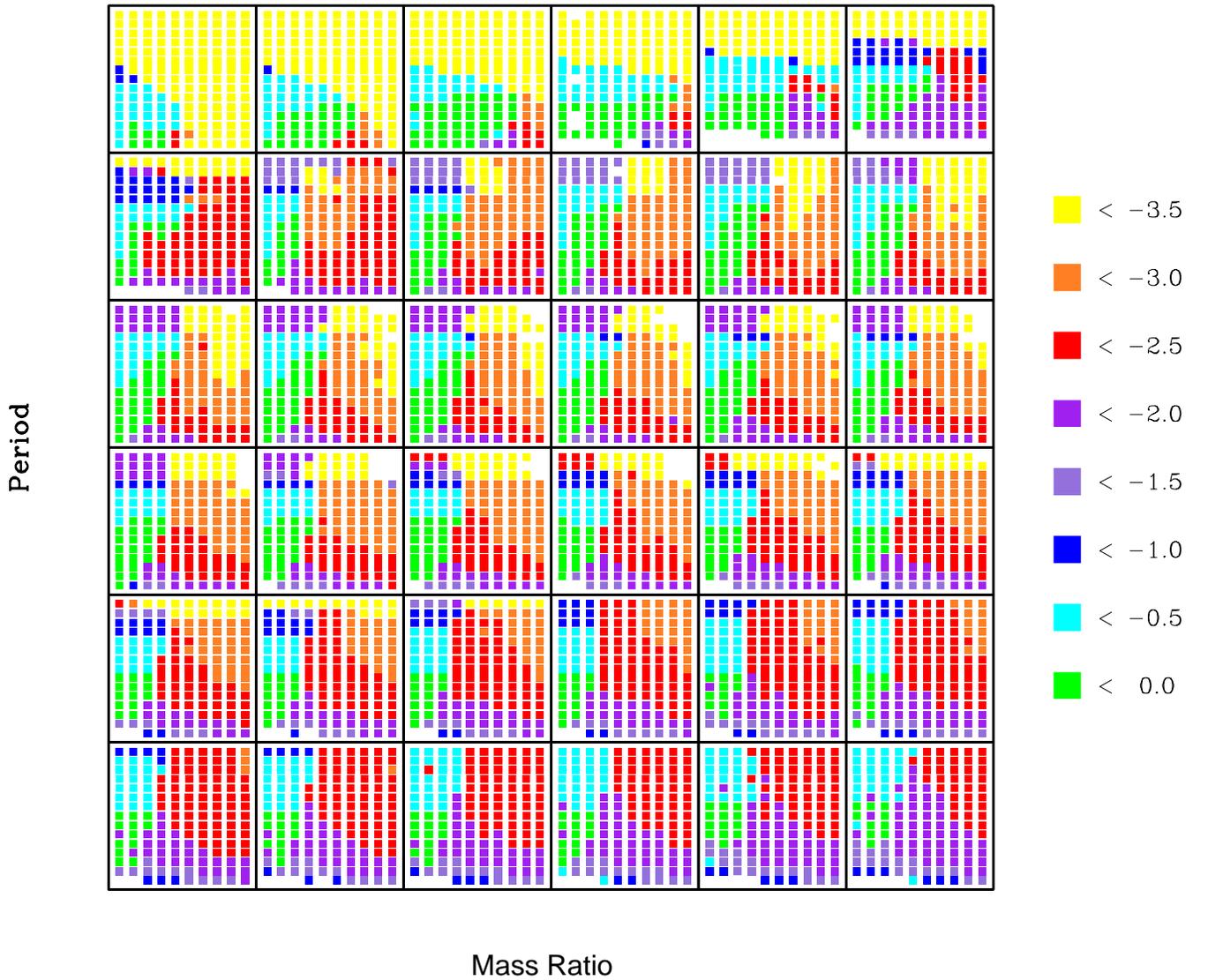}
\caption{The fraction of a given binary's lifetime time spent transferring
mass from $*1$ to $*2$, $f_{\rm RLOF}$. The ordinates are defined as in Figure~\ref{state}. The
blank squares are those binaries which failed to converge for more then a few
timesteps. The color-coded fractions are in logarithmic units.  The legend is
to be interpreted as follows: binaries with $\log f_{\rm RLOF} < -3.5$ are
shown as yellow squares, those with $-3.5 \leq f_{\rm RLOF} < -3.0$ as orange
squares, and so forth.}
\label{rlof}
\end{figure}

In Figures~\ref{mratio} and~\ref{rlof} we show, also in colour-coded form, the
following two properties of systems in our data cube: (i) the final mass ratio
of each system and (ii) the fraction of time spent as a semidetached system.
We define the state of the binary to be final when a) contact is reached, b)
reverse mass transfer begins, c) $*1$ has ignited carbon, or d) the binary is
detached and we believe reverse mass transfer to be imminent (ie. the function
$\log { \frac{R_{2}}{RL_{2}}(t) }$ will soon reach zero).  The time spent as a
semi-detached system has implications for the frequency of Algols in the
field.  We find that for a given primary mass the longest-lived Algols
originate from systems where mass transfer begins near the transition to the
HG (Late Case A to Early Case B or cases AL, AN, BL, BN) and with small to
moderate initial mass ratios.

\section{Comparison with Observed Systems}

Many observed binaries are semidetached (Algols), and one might hope 
that they could be matched by some of the above theoretical models during  
their stage of RLOF. However it has been clear for some time (Refsdal, Roth
\& Weigert 1974) that at least some systems (specifically, AS Eri) have such 
low angular momentum that they could hardly have started as detached systems 
of two zero-age main sequence stars of comparable mass. Furthermore there are 
some Algols of such low {\it total} mass (e.g. R CMa) that they also could 
hardly have started in such a configuration.

In an important paper, \citet{max96} identified 9 Algol systems 
for which they thought the observational data was of an unusually high quality.
They compared these with models computed by \citet{deg93}. The comparison
was not at all satisfactory, the theoretical models having luminosities
at least 20 times greater than the observed models. They also had substantially
longer periods. These discrepancies appear to be due to the following two features:
\renewcommand{\theenumi}{\roman{enumi}}
\renewcommand{\labelenumi}{(\theenumi)}
\begin{enumerate}
\item the theoretical models were all Case B
\item they were non-conservative, the assumption being made that 50\% of the
mass lost by $*1$ escaped to infinity, and 50\% was accreted by $*2$. The
escaping mass was assumed to remove the same specific angular momentum as
resided in the orbit of $*1$.  
\end{enumerate}

We feel that although the kind of non-conservation modeled by \citet{deg93}
may perhaps be appropriate for massive stars (O, and even early B), where
radiation pressure may be an important agent in mass loss, it is not
appropriate for mid-main sequence stars where, at least in single stars, very
little mass loss is normally observed. At the other end of the main sequence,
stellar winds are rather commonly observed, particularly in rapidly rotating
G/K/M dwarfs (and even more so in giants). These winds probably do not carry
off much mass (although see later), but they may be rich in angular momentum 
because of magnetic
linkage to the parent star. We therefore think that conservative models may be
reasonable for systems which are in the middle of the main sequence initially
(say B1 to G0), and where the loser has not yet evolved to the red-giant
region at spectra type $\sim$G or later. Following \citet{pop80} we refer to these systems
as `hot Algols'.  Unfortunately rather few of the \citet{max96} selection qualify
as hot Algols in this sense, although two (U CrB and AF Gem) are on the
border, with the cooler component having spectral type $\sim$G0. We have
therefore included a few more from the literature.  Our selection of hot
Algols is listed in Table 2, with references.

The observed parameters which we attempt to fit with our theoretical
models are the six independent quantities $\log P$, $\log M_1$, $\log q$,
$\log R_2$, $\log T_1$ and $\log T_2$. $R_1$ is not independent of these, 
since it is obtained from the \textit{assumption} that $*1$ fills its Roche
lobe, whose radius is determined by the first 3 parameters. $L_1$ and $L_2$
are similarly not independent of these 6 parameters. Our theoretical models
have four independent parameters, $\log P_0$, $\log M_{10}$, $\log q_0$ 
and age.

For each system in Tables 2 and 3 we give three rows. The first gives the
observational data from the literature, and the next the theoretical values
from our data cube which minimize $\chi^2$. The second row also includes the
best-fit age, in units of Myr. The third row gives the zero-age values for the
system which we infer from our best fit. We use mass-ratio $q$ because this is
usually obtained more directly from the observational data, whether
spectroscopic or photometric, than either $M_1$ or $M_2$.  We list
observational errors (when available) in the first row for all quantities, but
we list total errors (described below) in the second row only for those
quantities that we actually fit. 

\begin{figure}
\epsscale{0.5}
\plotone{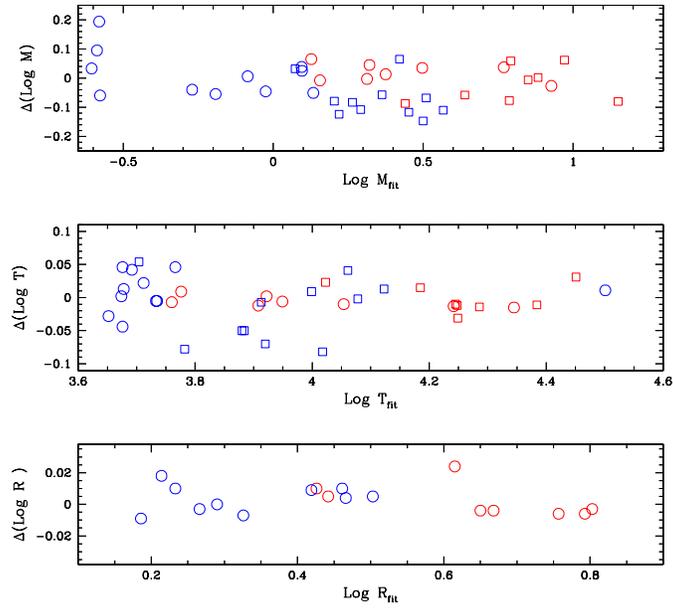}
\caption{The residuals of observed binary fits.  The hot Algols of Table 2 are shown
in red, the cool Algols of Table 3 are shown in blue.  The residuals of $*1$
are shown as circles, the residuals of $*2$ as squares.  The residuals are 
defined in the sense $\Delta = \rm{fit} - \rm{obs}$ .}
\label{fit}
\end{figure}

In fitting observed stars to theoretical models, a $\chi^2$ test
seems appropriate. However, we have to modify the standard test in order
to incorporate the fact that our theoretical models have an intrinsic
`graininess' because they have not been computed for a continuous
range of input parameters, but only at the grid-points in our data cube.
We therefore use a total error, $\sigma$, which is the sum in quadrature of
the observational error, $\sigma_{\rm obs}$, and a `theoretical error',
$\sigma_{\rm th}$, representing the intrinsic graininess.  For
$\log P,~\log M_1$ and $\log q$ we take $\sigma_{\rm th} = 0.05$, the initial
spacing of our grid.  For $\log R$ and $\log T$ we take the graininess
to be the difference in these parameters between adjacent ZAMS models from the
grid, centered on the mass of the observed binary.  For example, for an
observed star of mass $\log M = 1.02$ we take the theoretical error in the 
radius to be
\be
\sigma_{\rm th, R}(\log M = 1.02) = 
R_{\rm ZAMS}(\log M = 1.05) - R_{\rm ZAMS}(\log M = 1.00)~.
\ee
We can then look in our data cube for the minimum value of
\be
\chi^2 =\sum {\frac{({\rm obs-th})^2}{ \sigma_{\rm obs}^2+\sigma_{\rm th}^2}}~.
\ee
We find that the best fit point picked by minimizing this $\chi^2$ is
insensitive to the exact definition of $\sigma_{\rm th}$.  However, the
magnitude of $\chi^2_{\rm min}$ depends directly on $\sigma$, so we have
attempted a reasonable definition.  In Figure~\ref{fit} we present the
residuals to the fit for all Algols from both Tables 2 and 3.

The hot Algols of Table 2 have a mean $\chi^2$ of $\sim 3$.  Since there are 2
degrees of freedom (6 observed parameters less 4 theoretical parameters), this
value is rather more, but not enormously more, than is expected for a normal
distribution of errors.  The number of systems which we use is too small to
provide a really convincing confirmation or refutation.
The worst case, AF Gem, is very close to the lower temperature
limit, where we suppose {\it a priori} that conservation might break
down. If we reject AF Gem, we have a mean $\chi^2$ of just 2.

After AF Gem, the next worse cases are DM Per and $\lambda$ Tau.
Interestingly, both of these systems possess a close third body --
extraordinarily close in the case of $\lambda$ Tau. The latter system can be
seen to be problematic even without a detailed attempt at fitting. The angular
momentum of this system is seen to be quite low compared with a system of
comparable-mass stars at the same total mass, so that something like Case AS
is to be expected. But Case AS normally evolves into contact at a mass ratio
which is moderately small, roughly $\gtrsim0.4$ (Figure~\ref{mratio}), whereas
$\lambda$ Tau has quite a small present mass ratio of $0.27$. This suggests
that $\lambda$ Tau has lost some angular momentum, necessarily during its
slow, nuclear-timescale, RLOF rather than the comparable interval of detached
evolution before RLOF.  DM Per's problem is similar, though not so obvious
without a detailed attempt at fitting.

But for $\lambda$ Tau and DM Per, unlike most other hot Algols, there does in
fact exist a mechanism that should do just that. The third star in the
$\lambda$ Tau system \citep{fek82} is in such a close orbit ($33$d) that it
must influence the orbit of the eclipsing pair to a small but significant
extent, making its eccentricity fluctuate by $\sim 0.7\%$ on a timescale of
days \citep{kis98}. Tidal friction will tend to oppose this, but can only do
so by draining energy and angular momentum from the short-period orbit.
Conservation laws require the angular momentum lost by the inner orbit to go
to the outer orbit, but the energy loss leads to a net secular evolution, the
inner orbit shrinking while the outer widens.  This process was probably
negligible in the pre-RLOF state because the orbit would have been
substantially smaller than at present, at least if $q_0$ were not unusually
large. But it can now be significant as the stars are larger and the inner
orbit wider. Tidal friction should be capable of setting up a transient
equilibrium between nuclear evolution, leading to expansion of the inner
orbit, and tidal friction, leading to contraction \citep{kis98}. 

DM Per also has a third body in orbit.  The outer orbit is longer ($\sim
100$d) while the inner orbit is shorter, and so the process might be
thought less likely to be significant. But on the other hand the third body is
relatively much more massive, which may compensate to some extent.  

When we turn to a selection of cooler Algols (Table 3) we find significantly
larger $\chi^2$ for many systems. This, we believe, is consistent with the
view that they are less conservative, certainly of angular momentum (which is
fairly readily removed by magnetic braking on something like a nuclear
timescale), and perhaps also of mass. We do not normally think of stellar
winds from cool dwarfs and subgiants as being strong enough to remove
significant {\it mass}, and yet certain active (RS CVn) binaries show evidence
to the contrary. Both Z Her \citep{pop88} and to a lesser extent RW UMa
\citep{pop80,sca93} exhibit the phenomenon that the cooler, presumably more
evolved, subgiant is the {\it less} massive star, despite the fact that it
does not fill its Roche lobe. This suggests that mass loss by wind from the
cooler star is already on the nuclear timescale of the star.  

V1379 Aql \citep{jef97} is an example of a `post-Algol' binary: the low-mass
SDB component is presumably the remains of $*1$ after it has retreated within
its Roche lobe, and $*2$ has already evolved to the giant branch. We include it in 
our list of cool Algols as we believe the components to have been relatively cool
during its Algol phase.  The cool ZAMS temperatures which we derive support this
assumption. Even without detailed fitting, it is clear that the system is
problematic. For a period as short as 21d the SDB mass is rather large -- such
a core would seem to imply a period of 50 -- 100d. More intriguingly, the
orbit is very significantly eccentric: $e=0.09\pm 0.01$. Several radio pulsars
with WD companions are known with comparable period and with highly circular 
orbits -- e.g. 1855+09 \citep{ryb91} --
as expected following stable RLOF from a low-mass giant. Probably the least
far-fetched explanation of the eccentricity in V1379 Aql is the presence of a 
third body in a substantially inclined orbit; and this might also explain some
loss of angular momentum.

\section{Conclusions}

Case A RLOF, even when restricted to the classical `conservative' model, shows 
a rich variety of behaviour, which we feel is often not emphasised enough.
We identify 9 sub-classes, depending partly on whether the system evolves into
contact (in 5 different ways) or reaches reverse RLOF (in 4 different ways).
Further subdivision depends on the evolutionary states reached when contact or 
reverse RLOF occurs. We can expect even more subclasses when non-conservative
processes are modeled, as will certainly be necessary for extremes of 
high-temperature and low-temperature systems.

For all but one of our selection of observed hot Algols we find an acceptable 
$\chi^2$ when fitting the observed parameters to our library of conservative
Case A binary tracks.   It is encouraging to note that the worst outlier (AF
Gem) lies near the lower boundary of the temperature range in which we
expect the conservative assumption to hold.  The next largest $\chi^2$'s come
from two binaries with known third bodies, which may act to remove angular
momentum from the inner orbit.

Our selection of cool Algols shows significantly worse agreement between the
observed systems and the conservative theoretical tracks, suggesting the need
for more free parameters in the modelling, such as mass and angular momentum
loss.

This data set of conservative Case A tracks has uses beyond an indiviual
comparison of observed systems.  With an estimate of the initial mass function
and period distribution of binaries, it may be useful for population synthesis
studies or for creating close binary-inclusive isochrones for stellar
population studies.  We hope to make these tracks available in early 2001 on
the Institute of Geophysics and Planetary Physics web site
\url{http://www.llnl.gov/urp/IGPP}.

\acknowledgements

This work was undertaken as part of the DJEHUTY project, which is developing
a 3-D code to deal with hydrodynamical processes within stars, both single
and binary. Most of the end-points to which our Case A systems evolved can be
expected to have behaviour on a hydrodynamic timescale, and we hope to
investigate them further in the future. We are grateful to our DJEHUTY
colleagues for helpful discussions, and in particular to Don Dossa for helping
with the parallelisation of the code.  Work performed at LLNL is supported by
the DOE under contract W7405-ENG-48.  CAN is supported in part by a NPSC
Graduate Fellowship.


\clearpage

\clearpage
\begin{deluxetable}{cllll}
\tablewidth{0pc}
\tablecolumns{5}
\tabletypesize{\scriptsize}
\tablecaption{Summary of Binary Classification Scheme}
\tablehead{
\colhead{Case} &\colhead{Defining Equations} &\colhead{End State $*1$}
&\colhead{End State $*2$} &\colhead{End Geometry}
}
\startdata
AD & $\Mdot > M / t_{\rm dyn}$	&  M	&	M	& Contact \\
AR & $\Mdot > M / t_{\rm KH}$, $t_{\rm contact} - t_{\rm RLOF}(*1) < 0.10 \times t_{\rm MS}(*1)$ & M
& M & Contact \\
AS & $t_{\rm contact} - t_{\rm RLOF}(*1) > 0.10 \times t_{MS}(*1)$ &  M  & M & Contact \\
AE & $t_{\rm H}(*2)  < t_{\rm H}(*1)$	&  M	&	H	& Contact \\
AG & $t_{\rm H}(*1) < t_{\rm H}(*2)$   &  G	&	H,G	& Contact \\
AL  & $t_{\rm RLOF}(*2)<t_{\rm Cburn}(*1)$ & R,C
& H,G & RLOF $*2$ \\
AN  & $t_{\rm Cburn}(*1)<t_{\rm RLOF}(*2)$ &
SNe & M,H,G & Detached \\
\enddata
\end{deluxetable}

\clearpage
\begin{deluxetable}{lccccccccccccccccccc}  
\setlength{\tabcolsep}{0.06in}
\renewcommand{\arraystretch}{1.00}
\tablewidth{0pc}
\tablecolumns{21}
\tabletypesize{\scriptsize}
\tablecaption{Hot Algols: observed, best fit and ZAMS parameters}  
\rotate
\tablehead{  
\colhead{ }&\colhead{log}&\colhead{log}&\colhead{}
&\colhead{log}&\colhead{}&\colhead{log}&\colhead{}&\colhead{log}
&\colhead{}&\colhead{log}&\colhead{}&\colhead{log}&\colhead{}
&\colhead{log}&\colhead{}&\colhead{log}&\colhead{}&\colhead{}
&\colhead{} \cr
\colhead{Star}&\colhead{$P$}&\colhead{$M_{1}$}
&\colhead{$\sigma_{M_{1}}$}&\colhead{$q$}&\colhead{$\sigma_{q}$}
&\colhead{$T_{1}$}&\colhead{$\sigma_{T_{1}}$}&\colhead{$T_{2}$}&\colhead{$\sigma_{T_{2}}$}
&\colhead{$R_{1}$}&\colhead{$\sigma_{R_{1}}$}&\colhead{$R_{2}$}&\colhead{$\sigma_{R_{2}}$}
&\colhead{$L_{1}$}&\colhead{$\sigma_{L_{1}}$}&\colhead{$L_{2}$}&\colhead{$\sigma_{L_{2}}$}
&\colhead{Age}&\colhead{$\chi^{2}$}
}
\startdata  
   TT Aur$^9$ & 0.124 & 0.732 &  0.023 & -0.175 &  0.010 & 4.255 &  0.020 &  4.395 &  0.020 & 0.623 &  0.010 &  0.591 &  0.011 & 3.210 &  0.030 &  3.710 &  0.030 &\nodata & \nodata \\ 
\nodata & 0.149 & 0.769 &  0.055 & -0.201 &  0.051 & 4.242 &  0.035 &  4.384 &  0.033 & 0.650 &\nodata &  0.615 &  0.031 & 3.220 &\nodata &  3.720 & \nodata &    16 &  1.775 \\ 
\nodata & 0.119 & 0.950 & \nodata &  0.150 & \nodata & 4.369 & \nodata &  4.287 & \nodata & 0.581 & \nodata &  0.493 & \nodata & 3.593 & \nodata &  3.088 & \nodata &\nodata & \nodata \\ 
    U CrB$^4$ & 0.538 & 0.164 &  0.023 & -0.420 &  0.022 & 3.767 &  0.015 &  4.170 &  0.009 & 0.694 &  0.008 &  0.436 &  0.011 & 1.430 &  0.060 &  2.510 &  0.130 &\nodata & \nodata \\ 
\nodata & 0.547 & 0.158 &  0.055 & -0.481 &  0.055 & 3.761 &  0.049 &  4.187 &  0.031 & 0.703 &\nodata &  0.433 &  0.032 & 1.403 &\nodata &  2.567 & \nodata &   218 &  1.604 \\ 
\nodata & 0.235 & 0.550 & \nodata &  0.200 & \nodata & 4.137 & \nodata &  4.003 & \nodata & 0.341 & \nodata &  0.227 & \nodata & 2.185 & \nodata &  1.417 & \nodata &\nodata & \nodata \\ 
   AF Gem$^7$ & 0.095 & 0.063 &  0.015 & -0.466 &  0.010 & 3.767 &  0.010 &  4.000 &  0.020 & 0.365 &  0.007 &  0.417 &  0.010 & 0.750 &  0.030 &  1.780 &  0.090 &\nodata & \nodata \\ 
\nodata & 0.124 & 0.127 &  0.052 & -0.313 &  0.051 & 3.776 &  0.027 &  4.023 &  0.037 & 0.414 &\nodata &  0.426 &  0.032 & 0.883 &\nodata &  1.897 & \nodata &   635 & 11.394 \\ 
\nodata & 0.026 & 0.400 & \nodata &  0.200 & \nodata & 4.038 & \nodata &  3.878 & \nodata & 0.254 & \nodata &  0.169 & \nodata & 1.612 & \nodata &  0.803 & \nodata &\nodata & \nodata \\ 
    u Her$^5$ & 0.312 & 0.462 &  0.029 & -0.409 &  0.022 & 4.064 &  0.020 &  4.300 &  0.020 & 0.643 &  0.029 &  0.763 &  0.022 & 2.490 &  0.060 &  3.680 &  0.050 &\nodata & \nodata \\ 
\nodata & 0.330 & 0.497 &  0.058 & -0.386 &  0.055 & 4.054 &  0.039 &  4.286 &  0.033 & 0.673 &\nodata &  0.757 &  0.037 & 2.516 &\nodata &  3.612 & \nodata &    64 &  0.949 \\ 
\nodata & 0.120 & 0.800 & \nodata &  0.150 & \nodata & 4.287 & \nodata &  4.200 & \nodata & 0.493 & \nodata &  0.402 & \nodata & 3.088 & \nodata &  2.554 & \nodata &\nodata & \nodata \\ 
   DM Per$^6$ & 0.436 & 0.316 &  0.012 & -0.547 &  0.011 & 3.920 &  0.010 &  4.260 &  0.020 & 0.677 &  0.008 &  0.653 &  0.009 & 2.000 &  0.060 &  3.280 &  0.100 &\nodata & \nodata \\ 
\nodata & 0.488 & 0.313 &  0.052 & -0.474 &  0.051 & 3.908 &  0.039 &  4.248 &  0.034 & 0.715 &\nodata &  0.650 &  0.031 & 2.016 &\nodata &  3.243 & \nodata &   114 &  3.379 \\ 
\nodata & 0.187 & 0.700 & \nodata &  0.200 & \nodata & 4.229 & \nodata &  4.106 & \nodata & 0.432 & \nodata &  0.311 & \nodata & 2.735 & \nodata &  1.997 & \nodata &\nodata & \nodata \\ 
   V Pup$^8$ & 0.163 & 0.954 &  0.046 & -0.277 &  0.068 & 4.360 &  0.060 &  4.420 &  0.040 & 0.724 &  0.024 &  0.799 &  0.020 & 3.850 &  0.250 &  4.200 &  0.160 &\nodata & \nodata \\ 
\nodata & 0.185 & 0.927 &  0.068 & -0.223 &  0.085 & 4.345 &  0.065 &  4.451 &  0.045 & 0.724 &\nodata &  0.793 &  0.034 & 3.784 &\nodata &  4.344 & \nodata &    10 &  1.311 \\ 
\nodata & 0.117 & 1.100 & \nodata &  0.100 & \nodata & 4.444 & \nodata &  4.395 & \nodata & 0.668 & \nodata &  0.610 & \nodata & 4.065 & \nodata &  3.755 & \nodata &\nodata & \nodata \\ 
  $\lambda$Tau$^{2,3}$ & 0.597 & 0.276 &  0.009 & -0.578 &  0.007 & 3.920 &  0.030 &  4.280 &  0.040 & 0.724 &  0.016 &  0.806 &  0.007 & 2.110 &  \nodata &  3.690 &  \nodata &\nodata & \nodata \\ 
\nodata & 0.641 & 0.321 &  0.051 & -0.529 &  0.050 & 3.922 &  0.048 &  4.249 &  0.048 & 0.819 &\nodata &  0.803 &  0.030 & 2.281 &\nodata &  3.557 & \nodata &    99 &  2.918 \\ 
\nodata & 0.254 & 0.750 & \nodata &  0.200 & \nodata & 4.259 & \nodata &  4.138 & \nodata & 0.463 & \nodata &  0.341 & \nodata & 2.913 & \nodata &  2.185 & \nodata &\nodata & \nodata \\ 
    Z Vul$^{1,8}$ & 0.391 & 0.362 &  0.018 & -0.367 &  0.039 & 3.955 &  0.020 &  4.255 &  0.040 & 0.653 &  0.019 &  0.672 &  0.018 & 2.070 &  0.060 &  3.300 &  0.160 &\nodata & \nodata \\ 
\nodata & 0.387 & 0.375 &  0.053 & -0.417 &  0.063 & 3.949 &  0.041 &  4.245 &  0.049 & 0.670 &\nodata &  0.668 &  0.035 & 2.088 &\nodata &  3.268 & \nodata &   107 &  0.776 \\ 
\nodata & 0.137 & 0.700 & \nodata &  0.150 & \nodata & 4.229 & \nodata &  4.138 & \nodata & 0.432 & \nodata &  0.341 & \nodata & 2.735 & \nodata &  2.185 & \nodata &\nodata & \nodata \\
\enddata
\tablerefs{
(1) \citet{ces77};
(2) \citet{fek82};
(3) \citet{giu83}
(4) \citet{hei88};
(5) \citet{hil84};
(6) \citet{hil92}
(7) \citet{max95};
(8) \citet{pop80};
(9) \citet{pop91};
}
\end{deluxetable}

\begin{deluxetable}{lccccccccccccccccccc}  
\setlength{\tabcolsep}{0.06in}
\renewcommand{\arraystretch}{1.00}
\tablewidth{0pc}
\tablecolumns{21}
\tabletypesize{\scriptsize}
\tablecaption{Cool Algols: observed, best fit and ZAMS parameters}  
\rotate
\tablehead{  
\colhead{ }&\colhead{log}&\colhead{log}&\colhead{}
&\colhead{log}&\colhead{}&\colhead{log}&\colhead{}&\colhead{log}
&\colhead{}&\colhead{log}&\colhead{}&\colhead{log}&\colhead{}
&\colhead{log}&\colhead{}&\colhead{log}&\colhead{}&\colhead{}
&\colhead{} \cr
\colhead{Star}&\colhead{$P$}&\colhead{$M_{1}$}
&\colhead{$\sigma_{M_{1}}$}&\colhead{$q$}&\colhead{$\sigma_{q}$}
&\colhead{$T_{1}$}&\colhead{$\sigma_{T_{1}}$}&\colhead{$T_{2}$}&\colhead{$\sigma_{T_{2}}$}
&\colhead{$R_{1}$}&\colhead{$\sigma_{R_{1}}$}&\colhead{$R_{2}$}&\colhead{$\sigma_{R_{2}}$}
&\colhead{$L_{1}$}&\colhead{$\sigma_{L_{1}}$}&\colhead{$L_{2}$}&\colhead{$\sigma_{L_{2}}$}
&\colhead{Age}&\colhead{$\chi^{2}$}
}
\startdata  
    S Cnc$^{15,9}$ & 0.977 &-0.638 &  0.036 & -1.045 &  0.001 & 3.665 &  0.010 &  3.990 &  0.010 & 0.720 &  0.004 &  0.332 &  0.004 & 1.050 &  0.050 &  1.580 &  0.045 &\nodata & \nodata \\ 
\nodata & 1.033 &-0.605 &  0.062 & -0.897 &  0.050 & 3.678 &  0.039 &  3.920 &  0.036 & 0.773 &\nodata &  0.326 &  0.028 & 1.210 &\nodata &  1.283 & \nodata &  4182 & 14.447 \\ 
\nodata &-0.055 & 0.150 & \nodata &  0.250 & \nodata & 3.832 & \nodata &  3.680 & \nodata & 0.148 & \nodata & -0.144 & \nodata & 0.580 & \nodata & -0.614 & \nodata &\nodata & \nodata \\ 
    R CMa$^{12,13}$ & 0.055 &-0.775 &  0.049 & -0.801 &  0.029 & 3.630 &  0.030 &  3.860 &  0.030 & 0.025 &  0.028 &  0.196 &  0.027 &-0.410 &  0.160 &  0.760 &  0.180 &\nodata & \nodata \\ 
\nodata & 0.152 &-0.580 &  0.070 & -0.653 &  0.058 & 3.676 &  0.048 &  3.782 &  0.039 & 0.192 &\nodata &  0.214 &  0.077 & 0.041 &\nodata &  0.508 & \nodata & 19470 & 22.871 \\ 
\nodata &-0.319 & 0.000 & \nodata &  0.350 & \nodata & 3.751 & \nodata &  3.569 & \nodata &-0.050 & \nodata & -0.386 & \nodata &-0.143 & \nodata & -1.544 & \nodata &\nodata & \nodata \\ 
   RZ Cas$^5$ & 0.077 &-0.137 &  0.012 & -0.480 &  0.010 & 3.672 &  0.020 &  3.934 &  0.005 & 0.288 &  0.007 &  0.223 &  0.008 & 0.160 &  0.080 &  1.120 &  0.020 &\nodata & \nodata \\ 
\nodata & 0.109 &-0.192 &  0.051 & -0.412 &  0.051 & 3.674 &  0.043 &  3.884 &  0.036 & 0.296 &\nodata &  0.233 &  0.028 & 0.239 &\nodata &  0.954 & \nodata &  3971 &  5.382 \\ 
\nodata &-0.105 & 0.150 & \nodata &  0.200 & \nodata & 3.832 & \nodata &  3.718 & \nodata & 0.148 & \nodata & -0.102 & \nodata & 0.580 & \nodata & -0.379 & \nodata &\nodata & \nodata \\ 
   TV Cas$^4$ & 0.258 & 0.185 &  0.014 & -0.393 &  0.008 & 3.720 &  0.040 &  4.020 &  0.020 & 0.517 &  0.007 &  0.498 &  0.008 & 0.860 &  0.170 &  2.030 &  0.090 &\nodata & \nodata \\ 
\nodata & 0.265 & 0.134 &  0.052 & -0.376 &  0.051 & 3.766 &  0.060 &  4.061 &  0.037 & 0.509 &\nodata &  0.503 &  0.032 & 1.033 &\nodata &  2.202 & \nodata &   502 &  2.925 \\ 
\nodata & 0.097 & 0.450 & \nodata &  0.200 & \nodata & 4.072 & \nodata &  3.924 & \nodata & 0.282 & \nodata &  0.183 & \nodata & 1.806 & \nodata &  1.014 & \nodata &\nodata & \nodata \\ 
   AS Eri$^{1,10}$ & 0.426 &-0.682 &  0.018 & -0.968 &  0.012 & 3.720 &  0.030 &  3.930 &  0.030 & 0.340 &  0.023 &  0.196 &  0.016 & 0.470 &  \nodata &  1.060 &  \nodata &\nodata & \nodata \\ 
\nodata & 0.542 &-0.587 &  0.053 & -0.791 &  0.051 & 3.676 &  0.048 &  3.880 &  0.048 & 0.451 &\nodata &  0.186 &  0.029 & 0.558 &\nodata &  0.848 & \nodata &  3474 & 22.549 \\ 
\nodata &-0.005 & 0.150 & \nodata &  0.500 & \nodata & 3.832 & \nodata &  3.569 & \nodata & 0.148 & \nodata & -0.386 & \nodata & 0.580 & \nodata & -1.544 & \nodata &\nodata & \nodata \\ 
   TT Hya$^{9,15}$ & 0.842 &-0.229 &  0.132 & -0.646 &  0.002 & 3.680 &  0.010 &  3.990 &  0.010 & 0.769 &  0.029 &  0.290 &  0.028 & 1.200 &  0.080 &  1.500 &  0.070 &\nodata & \nodata \\ 
\nodata & 0.846 &-0.270 &  0.141 & -0.633 &  0.050 & 3.652 &  0.039 &  3.999 &  0.036 & 0.759 &\nodata &  0.290 &  0.040 & 1.080 &\nodata &  1.529 & \nodata &  2262 &  0.718 \\ 
\nodata & 0.226 & 0.200 & \nodata &  0.100 & \nodata & 3.879 & \nodata &  3.802 & \nodata & 0.167 & \nodata &  0.090 & \nodata & 0.803 & \nodata &  0.339 & \nodata &\nodata & \nodata \\ 
   AT Peg$^6$ & 0.059 & 0.021 &  0.012 & -0.325 &  0.008 & 3.690 &  0.017 &  3.920 &  0.005 & 0.332 &  0.006 &  0.270 &  0.006 & 0.380 &  0.070 &  1.190 &  0.030 &\nodata & \nodata \\ 
\nodata & 0.090 &-0.025 &  0.051 & -0.289 &  0.051 & 3.712 &  0.031 &  3.913 &  0.036 & 0.341 &\nodata &  0.266 &  0.027 & 0.483 &\nodata &  1.139 & \nodata &  2078 &  2.252 \\ 
\nodata & 0.054 & 0.250 & \nodata &  0.250 & \nodata & 3.924 & \nodata &  3.750 & \nodata & 0.182 & \nodata & -0.050 & \nodata & 1.014 & \nodata & -0.149 & \nodata &\nodata & \nodata \\ 
$\beta$\ Per$^{2,11}$ & 0.457 &-0.092 &  0.026 & -0.663 &  0.010 & 3.650 &  0.028 &  4.100 &  0.017 & 0.544 &  0.012 &  0.462 &  0.006 & 0.630 &  \nodata &  2.190 &  \nodata &\nodata & \nodata \\ 
\nodata & 0.554 &-0.085 &  0.056 & -0.537 &  0.051 & 3.692 &  0.047 &  4.018 &  0.036 & 0.626 &\nodata &  0.466 &  0.031 & 0.973 &\nodata &  1.959 & \nodata &  1135 & 15.972 \\ 
\nodata & 0.154 & 0.350 & \nodata &  0.200 & \nodata & 4.002 & \nodata &  3.831 & \nodata & 0.227 & \nodata &  0.150 & \nodata & 1.416 & \nodata &  0.578 & \nodata &\nodata & \nodata \\ 
   HU Tau$^8$ & 0.313 & 0.057 &  0.011 & -0.592 &  0.008 & 3.738 &  0.012 &  4.080 &  0.034 & 0.507 &  0.004 &  0.410 &  0.005 & 0.920 &  0.050 &  2.090 &  0.150 &\nodata & \nodata \\ 
\nodata & 0.414 & 0.095 &  0.051 & -0.405 &  0.051 & 3.733 &  0.028 &  4.078 &  0.045 & 0.594 &\nodata &  0.419 &  0.031 & 1.075 &\nodata &  2.102 & \nodata &   515 & 18.311 \\ 
\nodata & 0.247 & 0.450 & \nodata &  0.250 & \nodata & 4.072 & \nodata &  3.878 & \nodata & 0.282 & \nodata &  0.169 & \nodata & 1.806 & \nodata &  0.803 & \nodata &\nodata & \nodata \\ 
   TX Uma$^7$ & 0.486 & 0.072 &  0.014 & -0.606 &  0.010 & 3.740 &  0.016 &  4.110 &  0.010 & 0.627 &  0.007 &  0.451 &  0.006 & 1.170 &  0.065 &  2.300 &  0.040 &\nodata & \nodata \\ 
\nodata & 0.525 & 0.096 &  0.052 & -0.471 &  0.051 & 3.735 &  0.030 &  4.123 &  0.031 & 0.668 &\nodata &  0.461 &  0.031 & 1.229 &\nodata &  2.368 & \nodata &   373 &  8.113 \\ 
\nodata & 0.266 & 0.500 & \nodata &  0.250 & \nodata & 4.105 & \nodata &  3.924 & \nodata & 0.311 & \nodata &  0.183 & \nodata & 1.997 & \nodata &  1.015 & \nodata &\nodata & \nodata \\ 
V1379 Aql$^3$& 1.315 &-0.517 &  0.021 & -0.873 &  0.005 & 4.490 &  0.020 &  3.650 &  0.030 &-1.284 &  0.076 &  0.955 &  0.036 & 0.380 &  0.110 &  1.480 &  0.070 &\nodata & \nodata \\ 
\nodata & 1.372 &-0.577 &  0.054 & -0.998 &  0.050 & 4.501 &  0.043 &  3.704 &  0.047 &-0.997 &\nodata &  0.967 &  0.044 & 0.963 &\nodata &  1.702 & \nodata &  2585 & 10.138 \\ 
\nodata & 0.004 & 0.250 & \nodata &  0.200 & \nodata & 3.924 & \nodata &  3.775 & \nodata & 0.182 & \nodata &  0.016 & \nodata & 1.014 & \nodata &  0.085 & \nodata &\nodata & \nodata \\
\enddata
\tablerefs{
(1) \citet{ces78};
(2) \citet{van93};
(3) \citet{jef97};
(4) \citet{kha92};
(5) \citet{max94a};
(6) \citet{max94b};
(7) \citet{max95a};
(8) \citet{max95b};
(9) \citet{max96};
(10) \citet{pop73};
(11) \citet{ric88};
(12) \citet{sar96};
(13) \citet{tom85};
(14) \citet{van93}
}
\end{deluxetable}
 

\begin{thebibliography}{}
\bibitem[Cester et al.(1977)]{ces77} Cester, B., Fedel, B., Giuricin, G., Mardirossian, F. \& Pucillo, M. ~1977, \aap, 61, 469 
\bibitem[Cester et al.(1978)]{ces78} Cester, B., Fedel, B., Giuricin, G., Mardirossian, F. \& Mezetti, F. ~1978, \aap, 62, 291 
\bibitem[De Greve(1993)]{deg93} De Greve, J. P. 1993, A\&AS, 97, 527 
\bibitem[De Greve \& Packet(1990)]{deg90} De Greve, J. P. \& Packet, W. 
1990, \aap, 230, 97
\bibitem[Eggleton(1971)]{egg71} Eggleton, P. P. 1971, \mnras, 151, 351
\bibitem[Eggleton(1972)]{egg72} Eggleton, P. P. 1972, \mnras, 156, 361
\bibitem[Eggleton(2000)]{egg00} Eggleton, P. P. 2000, in the Brian Warner Symposium,
New Astronomy Reviews, 44, 111
\bibitem[Eggleton, Faulkner \& Flannery(1973)]{egg73} Eggleton, P. P., Faulkner, J. \&
Flannery, B. P. 1973, \aap, 23, 325
\bibitem[Fekel \& Tomkin(1982)]{fek82} Fekel, F. C. \& Tomkin, J. 1982, 
\apj, 263, 289 
\bibitem[Giuricin, Mardirossian \& Mezzetti(1983)]{giu83} Giuricin, G., Mardirossian, F. \& Mezzetti, M. 1983, \apjs, 52, 35 
\bibitem[Heintze \& van Gent(1988)]{hei88} Heintze, J. R. W.
\& van Gent, R. H. 1988, in `Algols', ed. Batten, A. H. Kluwer AP, p264 
\bibitem[Hilditch(1984)]{hil84} Hilditch, R. W. 1984, \mnras, 211, 943 
\bibitem[Hilditch, Hill \& Khalesseh(1992)]{hil92} Hilditch, R. W.,  Hill, G.
\& Khalesseh, B. 1992, \mnras, 254, 82 
\bibitem[Jeffery \& Simon(1997)]{jef97} Jeffery, C. S. \& Simon, T. 1997,
\mnras, 286, 487 
\bibitem[Khalesseh \& Hill(1992)]{kha92} Khalesseh, B. \& Hill, G. 1992,
\aap, 257, 199 
\bibitem[Kiseleva, Eggleton \& Mikkola(1998)]{kis98} Kiseleva, L. G., Eggleton,
P. P. \& Mikkola, S. 1998, \mnras, 300, 292 
\bibitem[Maxted \& Hilditch(1995)]{max95} Maxted, P. F. L. \& Hilditch, R. W. 
1995, \aap, 301, 149 
\bibitem[Maxted \& Hilditch(1996)]{max96} Maxted, P. F. L. \& Hilditch, R. W.
1996 \aap, 311, 567 
\bibitem[Maxted, Hill \& Hilditch(1994a)]{max94a} Maxted, P. F. L., 
Hill, G. \& Hilditch, R. W. 1994a, \aap, 282, 821 
\bibitem[Maxted, Hill, \& Hilditch(1994b)]{max94b} Maxted, P. F. L., 
Hill, G. \& Hilditch, R. W. 1994b, \aap, 285, 535 
\bibitem[Maxted, Hill, \& Hilditch(1995a)]{max95a} Maxted, P. F. L., 
Hill, G. \& Hilditch, R. W. 1995a, \aap, 301, 135 
\bibitem[Maxted, Hill, \& Hilditch (1995b)]{max95b} Maxted, P. F. L.,
Hill, G. \& Hilditch, R. W. 1995b, \aap, 301, 141 
\bibitem[Paczy{\'n}ski(1976)]{pac76} Paczy{\'n}ski, B. 1976,
Structure and Evolution of Close Binary Systems; IAU Symp. 73,
eds P. Eggleton, S. Mitton, and J. Whelan. Reidel: Dordrecht. p.75
\bibitem[Pols(1994)]{pol94} Pols, O. R. 1994, A\&A, 290, 119 
\bibitem[Pols et al.(1997)]{pol97} Pols, O. R., Tout, C. A., 
Schr{\"o}der, K.-P., Eggleton, P. P. \& Manners, J.~1997, \mnras, 289, 869
\bibitem[Popper(1973)]{pop73} Popper, D. M. 1973, \apj, 185, 265 
\bibitem[Popper(1980)]{pop80} Popper, D. M. 1980, \araa, 18, 115
\bibitem[Popper(1988)]{pop88} Popper, D. M. 1988, \aj, 96, 1040
\bibitem[Popper \& Hill(1991)]{pop91} Popper, D. M. \& Hill, G. 
1991, \aj, 101, 600  
\bibitem[Richards, Mochnacki, \& Bolton(1988)]{ric88} Richards, M. T.,
Mochnacki, S. W. \& Bolton, C. T. 1988, \aj, 96, 326 
\bibitem[Ryba \& Taylor(1991)]{ryb91} Ryba, M. F. \& Taylor, J. H. 1991, 
ApJ, 371, 739 
\bibitem[Sarma, Vivekananda Rao \& Abhyankar(1996)]{sar96} 
Sarma, M. B. K., Vivekananda Rao, P. \& Abhyankar, K. D. 1996, 
\apj, 458, 371 
\bibitem[Scaltriti et al.(1993)]{sca93} Scaltriti, F., Busso, M.,
Ferrari-Toniolo, M., Origlia, L., Persi, P., Robberto, M. \& Silvestro, G. 
1993, \mnras, 264, 5
\bibitem[Schr{\"o}der, Pols \& Eggleton(1997)]{sch97} Schr{\"o}der, K.-P.,
Pols, O. R. \& Eggleton, P. P. 1997, \mnras, 285, 696
\bibitem[Schwarzschild \& H{\"a}rm(1958)]{sch58} Schwarzschild, M. \& H{\"a}rm, 
R., 1958, \apj, 128, 348
\bibitem[Tomkin(1985)]{tom85} Tomkin, J. 1985, \apj, 297, 250 
\bibitem[Tout \& Eggleton(1988)]{tou88} Tout, C. A. \& Eggleton, P. P. 1988,
\apj, 334, 357
\bibitem[Tout, et al.(1996)]{tou96} Tout,  C. A., Pols, O. R., Han, Zh. \&
Eggleton, P. P. 1996, \mnras, 281, 257
\bibitem[Van Hamme \& Wilson(1993)]{van93} Van Hamme, W. \& Wilson,
R. E. 1993, \mnras, 262, 220 

\end{thebibliography}
\end{document}